\documentclass[12pt]{article}

\usepackage{mathtools,amsfonts,amssymb}
\usepackage{array}
\usepackage{epsfig}
\usepackage{graphicx}
\usepackage[centertableaux,boxsize=1.1em]{ytableau}
\usepackage{hyperref}
\usepackage{bm}
\usepackage{fancyhdr}
\usepackage[margin=1in]{geometry}
\usepackage{booktabs}
\usepackage[shortlabels,inline]{enumitem}
\usepackage[nameinlink,capitalize]{cleveref}

\DeclareMathOperator{\Adj}{Adj}
\DeclareMathOperator{\SU}{SU}
\DeclareMathOperator{\U}{U}
\newcommand*{\cN}{\mathcal{N}}
\newcommand*{\bP}{\mathbb{P}}
\newcommand*{\Z}{\mathbb{Z}}
\newcommand*{\yng}[1]{\ydiagram{#1}}
\newcommand*{\syng}[1]{\text{\small$\ydiagram{#1}$}}
\renewcommand{\bar}[1]{\ensuremath\overline{#1}}

\newcommand*{\SMn}{\ensuremath\SU(3) \times \SU(2) \times \U(1)}
\newcommand*{\SM}{\ensuremath(\SU(3) \times \SU(2) \times \U(1)) / \Z_6}
\newcommand*{\SMuh}{\ensuremath\SU(4) \times \SU(3) \times \SU(2)}
\newcommand*{\PS}{\ensuremath(\SU(4) \times \SU(2) \times \SU(2)) / \Z_2}

\makeatletter
\def\blfootnote{\gdef\@thefnmark{}\@footnotetext}
\makeatother

\title{Generic construction of the Standard Model gauge group and matter
representations in F-theory}
\author{Washington Taylor and Andrew P. Turner \\[0.9em]
\small\textit{Center for Theoretical Physics, Department of Physics} \\
\small\textit{Massachusetts Institute of Technology} \\%
\small\textit{77 Massachusetts Avenue} \\%
\small\textit{Cambridge, MA 02139, USA}%
}
\date{\today}

\begin{document}
\maketitle
\thispagestyle{fancy}

\vspace{-2em}
\begin{abstract}
We describe general classes of 6D and 4D F-theory models with gauge group
$\SM$.  We prove that this set of constructions gives all possible consistent
6D supergravity theories with no tensor multiplets having this gauge group and
the corresponding generic matter representations, which include those of the
MSSM. We expect, though do not prove, that these models are similarly generic
for 6D theories with tensor multiplets and for 4D $\cN = 1$ supergravity
theories. The largest class of these constructions comes from deforming an
underlying geometry with gauge symmetry $\SMuh$.
\end{abstract}

\blfootnote{
\texttt{wati} at \texttt{mit.edu},
\texttt{apturner} at \texttt{mit.edu}
}
\vspace{-1.2em}

\tableofcontents

%%%%%%%%%%%%%%%%%%%%%%%%%%%%%%%%%%%%%%%%%%%%%%%%%%%%%%%%%%%%%%%%%%%%%%%%%%%%%%
%%%%%%%%%%%%%%%%%%%%%%%%%%%%%%%%%%%%%%%%%%%%%%%%%%%%%%%%%%%%%%%%%%%%%%%%%%%%%%
%%%%%%%%%%%%%%%%%%%%%%%%%%%%%%%%%%%%%%%%%%%%%%%%%%%%%%%%%%%%%%%%%%%%%%%%%%%%%%
\section{Introduction}\label{sec:intro}

Since the early days of string theory, ongoing efforts have been made to
construct vacuum solutions that match some or all of the features of the
observed Standard Model of particle physics.  A tremendous range of such models
has been produced over the years, using a variety of geometric and algebraic
techniques, and there is increasing evidence that string theory can describe a
large number of solutions with Standard Model structure (see, e.g.,
\cite{AndersonEtalScan,ConstantinHeLukasCountingSM}).  In recent years, F-theory
\cite{VafaF-theory,MorrisonVafaI,MorrisonVafaII} has emerged as a powerful
framework for studying the global structure of the space of string vacua.  In
six dimensions, the connected space of F-theory vacua reproduces much of the
structure of the moduli space of consistent 6D $\cN = (1, 0)$ supergravity
theories, and under certain restrictions on the number of tensor multiplets,
the gauge group, and the matter representation content, one can prove that
F-theory describes all such consistent 6D theories. In this note, we use the
strong connection between F-theory and 6D supergravity theories to determine
the general form of F-theory models that reproduce the Standard Model gauge
group and a set of ``generic'' matter representations that includes those of
the Standard Model. These models have a natural realization in 4D as well,
giving a large, and perhaps fairly complete, classification of 4D $\cN = 1$
F-theory models in which the Standard Model arises through a tuned geometry.

There are a number of ways in which the Standard Model could, in principle,
arise in the framework of F-theory.  One approach, which has been studied
extensively, is to realize the gauge group of the Standard Model through flux
breaking of a (geometrically tuned) GUT such as $\SU(5)$
\cite{DonagiWijnholtGUTs,DonagiWijnholtModelBuilding,
BeasleyHeckmanVafaI,BeasleyHeckmanVafaII} (see \cite{WeigandTASI} for a
review).  A second approach is based on the idea that all or part of the
Standard Model gauge group may be a generic (i.e., supersymmetrically
``non-Higgsable'' \cite{MorrisonTaylorClusters,MorrisonTaylor4DClusters})
feature in the chosen geometry, either directly as explored in
\cite{GrassiHalversonShanesonTaylor} or through flux breaking of a larger
non-Higgsable GUT group such as $E_8$ \cite{TaylorWangVacua,TianWangEString}.
A third approach, and the one we focus on here, is that the gauge group may
itself be a tuned feature of the geometry. This approach was used, for
example, in \cite{LinWeigandSM,CveticEtAlThreeParam,LinWeigandG4,
CveticEtAlMSSMZ2}; more recently, Cveti\v{c}, Halverson, Lin, Liu, and Tian
described a large set of models where the Standard Model gauge group and
matter content are tuned over a weak Fano base \cite{CveticEtAlQuadrillion}.
The work in this note gives a general and comprehensive picture of the generic
classes of models that realize this gauge group and corresponding matter
representations through a tuned geometry in the F-theory context; the models
studied in \cite{CveticEtAlQuadrillion} are special cases of those described
here.

A key concept for our analysis is the notion of ``generic'' matter types for a
given gauge group.  As described in \cite{TaylorTurnerGeneric}, this idea can
be made rigorous in the context of 6D supergravity by defining the generic
matter representations for a given gauge group as those that appear on the
moduli space branch of largest dimension, for fixed and small anomaly
coefficients.  In simple cases, the number of generic matter fields defined in
this way matches the number of anomaly constraints, so there is a unique
solution for generic matter content for given anomaly coefficients.
Furthermore, the generic matter types defined in this way in six dimensions
match naturally to the simplest singularity types encountered in F-theory
constructions of the given gauge groups and the underlying geometric moduli of
F-theory models, so from the point of view of F-theory, this notion of
genericity naturally extends to four dimensional theories.

With this definition, the matter content of the Standard Model is not generic
for the gauge group $\SMn$, but it is generic for the group with global
structure $\SM$.  Thus, in this note we focus on the latter gauge group.  We
proceed by first finding a complete solution of the 6D anomaly equations for
the generic matter content of a theory with global gauge group $\SM$
(\cref{sec:6D}), and then describing how these models are realized in 6D
F-theory and generalizing the construction to 4D F-theory
(\cref{sec:F-theory}).  In other forthcoming work, we will describe more details
of this framework, both for the gauge group $\SU(5) \times \U(1)$ and its
quotient, and for the Standard Model gauge group.

%%%%%%%%%%%%%%%%%%%%%%%%%%%%%%%%%%%%%%%%%%%%%%%%%%%%%%%%%%%%%%%%%%%%%%%%%%%%%%
%%%%%%%%%%%%%%%%%%%%%%%%%%%%%%%%%%%%%%%%%%%%%%%%%%%%%%%%%%%%%%%%%%%%%%%%%%%%%%
%%%%%%%%%%%%%%%%%%%%%%%%%%%%%%%%%%%%%%%%%%%%%%%%%%%%%%%%%%%%%%%%%%%%%%%%%%%%%%
\section{SM gauge group and matter in 6D supergravity}\label{sec:6D}

%%%%%%%%%%%%%%%%%%%%%%%%%%%%%%%%%%%%%%%%%%%%%%%%%%%%%%%%%%%%%%%%%%%%%%%%%%%%%%
%%%%%%%%%%%%%%%%%%%%%%%%%%%%%%%%%%%%%%%%%%%%%%%%%%%%%%%%%%%%%%%%%%%%%%%%%%%%%%
\subsection{Solutions of 6D anomaly equations}\label{sec:6D-AC}

As described in \cite{TaylorTurnerGeneric}, for the gauge group $\SM$ there are
ten generic charged matter fields, listed in \cref{tab:generic}. The 6D gauge,
gravitational, and mixed anomaly equations \cite{GreenSchwarzWest6DAnom,
SagnottiGS} give ten constraints on the multiplicities of these fields; the
left-hand sides of the anomaly equations are inner products $a \cdot b, b \cdot
b$ of the anomaly coefficients $a, b_3, b_2, \tilde{b}$ associated with gravity
and the three gauge factors, respectively (using the notation of
\cite{KumarMorrisonTaylorGlobalAspects,TaylorTurnerGeneric}).  The ten anomaly
equations restricted to generic matter fields are an invertible system of
linear equations, so we can simply write the general solution of these
equations for the multiplicities of each of the fields in terms of the various
anomaly coefficient products, as listed in \cref{tab:generic}. We find it
convenient to use the following relations to define the quantities $\beta, X,
Y$:
\begin{equation}
\begin{aligned}
\tilde{b} & = \frac{4}{3} b_3 + \frac{3}{2} b_2 + 2 \beta\,, \\
X &= -8 a - 4 b_3 - 3 b_2 - 2 \beta\,, \\
Y &= a + b_3 + b_2 + \beta\,.
\end{aligned}
\end{equation}

\begin{table}[h!]
\centering

\[\begin{array}{ccc}\toprule
\text{Generic Matter} & \text{Multiplicity} & \text{MSSM Multiplet} \\ \midrule
\left(\yng{1}, \yng{1}\right)_{\mathrlap{1 / 6}} & b_3 \cdot b_2 &
    \bm{Q} \\[0.8em]
\left(\yng{1}, \bm{1}\right)_{\mathrlap{2 / 3}} & b_3\cdot (\beta - 2 a) &
    \bar{\bm{U}^c} \\[0.8em]
\left(\yng{1}, \bm{1}\right)_{\mathrlap{-1 / 3}} & b_3\cdot X &
    \bar{\bm{D}^c} \\[0.8em]
\left(\yng{1}, \bm{1}\right)_{\mathrlap{-4 / 3}} & b_3 \cdot Y & \\[0.8em]
\left(\bm{1}, \yng{1}\right)_{\mathrlap{1 / 2}} & b_2\cdot (X + \beta - a) &
    \bar{\bm{L}}, \bm{H}_u, \bar{\bm{H}_d}  \\[0.8em]
\left(\bm{1}, \yng{1}\right)_{\mathrlap{3 / 2}} & b_2\cdot Y & \\[0.8em]
(\bm{1}, \bm{1})_{\mathrlap{1}} & (b_3+ b_2+2 \beta) \cdot X - a \cdot b_2 &
    \bm{E}^c \\[0.8em]
(\bm{1}, \bm{1})_{\mathrlap{2}} & \beta \cdot Y & \\[0.8em]
(\Adj, \bm{1})_{\mathrlap{0}} & 1 + b_3 \cdot (b_3 + a) / 2 & \\[0.8em]
(\bm{1}, \Adj)_{\mathrlap{0}} & 1 + b_2 \cdot (b_2 + a) / 2 & \\ \bottomrule
\end{array}\]

\caption{Generic matter representations (not including conjugates) charged
under the gauge group $\SM$, which include all the charged MSSM
multiplets. Multiplicities given are for the generic matter solution of the 6D
anomaly equations, with $\beta, X, Y$ defined in the text. The dot product
between anomaly coefficients uses the signature-$(1, T)$ inner product defined
in a 6D supergravity theory with $T$ tensor fields.}
\label{tab:generic}
\end{table}

%%%%%%%%%%%%%%%%%%%%%%%%%%%%%%%%%%%%%%%%%%%%%%%%%%%%%%%%%%%%%%%%%%%%%%%%%%%%%%
%%%%%%%%%%%%%%%%%%%%%%%%%%%%%%%%%%%%%%%%%%%%%%%%%%%%%%%%%%%%%%%%%%%%%%%%%%%%%%
\subsection{$T = 0$ and two classes of solutions}\label{sec:6D-T=0}

We now analyze the resulting matter spectra, motivated by the case with no
tensor multiplets ($T = 0$), for which the anomaly coefficients $b_2, b_3$ are
integers, $a = -3$, and the inner product is simply multiplication.  For
nontrivial gauge groups $\SU(3)$ and $\SU(2)$ with good kinetic terms we must
have
\begin{equation}
\label{eq:bPositive}
b_2, b_3 > 0\,.
\end{equation}
At $T = 0$, this implies that $X$ and $Y$ must be non-negative for non-negative
spectra, so we have
\begin{equation}
\label{eq:X-constraint}
 4 b_3 + 3 b_2 + 2 \beta \le -8 a\,.
\end{equation}
and
\begin{equation}
\label{eq:Y-constraint}
b_3 + b_2 + \beta \ge -a
\end{equation}
The total number of $T = 0$ models satisfying these constraints is 98.

We can now define two classes of solutions, which cover all $T = 0$ solutions:
\begin{enumerate}[label=(\Alph*)]
\item\label{item:beta>=0}
$\beta \ge 0$.  This gives a three-parameter family of models, parametrized by
$b_3, b_2, \beta$ subject to the constraints
\labelcref{eq:X-constraint,eq:Y-constraint}; there are 71 such models for $T =
0$. These models precisely correspond to the Higgsing of a 6D supergravity
model with gauge group $\SMuh$ on a triplet of bifundamental fields (one
between each pair of gauge factors) in such a way as to preserve the symmetry
$\SM$, where the gauge factors have the anomaly coefficients $B_4 = b_3, B_3 =
b_2, B_2 =
\beta$.

\item\label{item:Y=0}
$Y = 0$.  In this case we can have $\beta < 0$, but we must have
\begin{equation}
2 a \le \beta = -a - b_3 - b_2\,.
\end{equation}
This gives a two-parameter family of models parametrized by $b_3, b_2$; there
are 30 such models for $T = 0$. Three of these models also fit into class
\labelcref{item:beta>=0}. All but three of the models in class
\labelcref{item:Y=0} correspond to Pati--Salam models, in which we Higgs a 6D
supergravity model with gauge group $\PS$ on a pair of bifundamental fields,
where the gauge factors have anomaly coefficients $B_4 = b_3, B_2 = b_2, B_2' =
-4 a - 2 b_3 - b_2$. The three exceptional cases without Pati--Salam
descriptions are those that would have $B_2' < 0$, and also do not have a
description in class \labelcref{item:beta>=0}. Note that the class
\labelcref{item:Y=0} model with $b_3 = b_2 = -a$ can also arise from Higgsing
an $\SU(5)$ theory with $B_5 = -a$ on an adjoint representation.
\end{enumerate}

The classes \labelcref{item:beta>=0} and \labelcref{item:Y=0} of models with
$\beta \ge 0$ and $Y = 0$, respectively, generalize naturally to theories with
arbitrary numbers of tensor multiplets $T > 0$, where $\beta \ge 0$ now
represents the condition that $\beta$ lies in the positivity cone, and the
condition \labelcref{eq:bPositive} becomes the statement that $b_2, b_3$ must
be nonzero and lie in the positivity cone. Models of class
\labelcref{item:beta>=0} that satisfy condition \labelcref{eq:X-constraint} can
again be interpreted as Higgsed $\SMuh$ models; similarly, a subset of models
of class \labelcref{item:Y=0} can be realized as Higgsed Pati--Salam models.

We note, however, that the story is more subtle when $T > 0$ since, for
example, $b_3 \cdot X \ge 0$ with $b_3$ in the positivity cone of the theory
does not necessarily imply that $X$ is in the positivity cone; similarly, when
$T > 0$ one can have vectors $\beta \cdot Y \ge 0$ with $\beta$ outside the
positivity cone and $Y$ nonzero.  In general, constraints on inner products are
weaker than the corresponding constraints on the factors, as discussed in
\cite{TaylorTurnerGeneric} in the context of the simple cases of theories with
$\U(1)$ and $\SU(2)$ gauge groups.

Note that both classes of models in general contain hypermultiplet matter in
the adjoint representations of the gauge factors $\SU(3), \SU(2)$, and the
models of class \labelcref{item:beta>=0} also generally have three other types
of matter fields not found in the MSSM.

Note also that as the gauge group becomes more complicated, non-generic matter
representations may become more relevant.  In particular, for the gauge group
$\SM$, the representation $\left(\syng{1}, \syng{1}\right)_{-5 / 6}$
is non-generic, but appears in theories coming from Higgsing $\SU(5)$ models
where the $\SU(5)$ starts with more than one adjoint (i.e., $B_5 > -a$).  While
in general the anomaly coefficients must be larger to realize models that
contain such a non-generic representation, such models may also be of interest
to study.

In summary, we have defined two classes of models with gauge group $\SM$ and
generic matter, which in the $T = 0$ case describe all solutions of the 6D
supergravity anomaly equations.

%%%%%%%%%%%%%%%%%%%%%%%%%%%%%%%%%%%%%%%%%%%%%%%%%%%%%%%%%%%%%%%%%%%%%%%%%%%%%%
%%%%%%%%%%%%%%%%%%%%%%%%%%%%%%%%%%%%%%%%%%%%%%%%%%%%%%%%%%%%%%%%%%%%%%%%%%%%%%
%%%%%%%%%%%%%%%%%%%%%%%%%%%%%%%%%%%%%%%%%%%%%%%%%%%%%%%%%%%%%%%%%%%%%%%%%%%%%%
\section{F-theory construction of the Standard Model gauge group and
generic matter representations}\label{sec:F-theory}

%%%%%%%%%%%%%%%%%%%%%%%%%%%%%%%%%%%%%%%%%%%%%%%%%%%%%%%%%%%%%%%%%%%%%%%%%%%%%%
%%%%%%%%%%%%%%%%%%%%%%%%%%%%%%%%%%%%%%%%%%%%%%%%%%%%%%%%%%%%%%%%%%%%%%%%%%%%%%
\subsection{F-theory constructions at $T = 0$}\label{sec:F-theory-T=0}

We now consider how the supergravity models with gauge group $\SM$ and generic
matter spectra given by \cref{tab:generic} can be realized in F-theory.  We
begin by considering the 6D $T = 0$ models.

As stated above, there are 71 combinations of $b_2, b_3 > 0, \beta \ge 0$ that
satisfy the constraints \labelcref{eq:X-constraint,eq:Y-constraint}.  It is
straightforward to confirm that these are precisely the combinations of divisor
classes on $\bP^2$ that allow for a Tate tuning
\cite{BershadskyEtAlSingularities,KatzEtAlTate} of $\SMuh$ on the corresponding
divisor classes $B_4 = b_3, B_3 = b_2, B_2 = \beta$. Since there are no
obstructions to flat directions in the 6D theory, the Higgsing to the desired
$\SM$ group can always be carried out. Thus, there is an F-theory construction
of all the models in class \labelcref{item:beta>=0}.

The models of class \labelcref{item:Y=0} precisely correspond to the
construction of F-theory models with toric fiber $F_{11}$ described in
\cite{KleversEtAlToric}; the global structure of such F-theory models with a
gauge group that is a quotient of a product of nonabelian and abelian factors
was analyzed in \cite{GrimmKapferKleversArithmetic,CveticLinU1}.  The matter
multiplicities computed in \cite{KleversEtAlToric} for models with the
$F_{11}$ fiber can be matched to the multiplicities of \cref{tab:generic}
through $b_3 = \mathcal{S}_9, b_2 = \mathcal{S}_7 - \mathcal{S}_9 - K_B$.  The
allowed values for the F-theory models precisely match those of the
supergravity models, giving 30 possible distinct models, so again there is an
F-theory construction of all the models in class \labelcref{item:Y=0}. For all
but three of the models of class \labelcref{item:Y=0}, there is also a
construction through the Higgsing of a Pati--Salam model, again starting with a
Tate tuning. This Higgsing can be related in the toric language to deformations
that give fiber $F_{11}$ by removing a ray from the fiber $F_{13}$, which as
pointed out in \cite{KleversEtAlToric} generally describes a Pati--Salam model.

We have thus shown that there is an F-theory realization for all possible
consistent 6D supergravity theories with $T = 0$, gauge group $\SM$, and matter
in the set of generic matter representations for this group as tabulated in
\cref{tab:generic}.

%%%%%%%%%%%%%%%%%%%%%%%%%%%%%%%%%%%%%%%%%%%%%%%%%%%%%%%%%%%%%%%%%%%%%%%%%%%%%%
%%%%%%%%%%%%%%%%%%%%%%%%%%%%%%%%%%%%%%%%%%%%%%%%%%%%%%%%%%%%%%%%%%%%%%%%%%%%%%
\subsection{Larger $T$}\label{sec:F-theory-T>0}

More generally, the classes of F-theory models used to realize the 6D
supergravity models with gauge group $\SM$ and generic matter can be
constructed over a general F-theory base $B_2$, giving a theory
with $T = h^{1, 1}(B_2) + 1$ tensor multiplets. It is natural to speculate that
such models, for class \labelcref{item:beta>=0} comprising the set of models
arising from Higgsing a theory with gauge group $\SMuh$ on three
bifundamentals, and for class \labelcref{item:Y=0} comprising the set of models
with fiber $F_{11}$ as described in \cite{KleversEtAlToric}, may give generic
constructions at arbitrary $T$ of the set of possible F-theory models with
gauge group $\SM$ and generic matter. For the models of class
\labelcref{item:Y=0}, indeed the construction of \cite{KleversEtAlToric}
essentially gives a parametrized Weierstrass model, analogous to a Tate model
for tuning an $\SU(N)$ theory or the Morrison--Park \cite{MorrisonParkU1} model
for tuning a $\U(1)$ theory with generic matter.  While we leave the explicit
construction of a Weierstrass model for the general theory of class
\labelcref{item:beta>=0} to further work \cite{RaghuramTaylorTurner321W},
it is clear that for an arbitrary number of tensor multiplets and a general
F-theory base with canonical class $K$, the constraints
\labelcref{eq:X-constraint,eq:Y-constraint}, where $a = K$, are precisely the
constraints on the existence of a Tate model for the necessary $\SMuh$ theory
(in the absence of other gauge factors).

There is an additional subtlety that at larger values of $T$, there can be
non-generic constructions that nonetheless only give rise to generic matter; we
do not discuss such constructions here, but they may provide additional exotic
models that nonetheless only contain the generic $\SM$ matter.

All of these constructions can be carried out over an arbitrary weak Fano
base, as discussed for certain models of class \labelcref{item:Y=0} in
\cite{CveticEtAlQuadrillion}. Note however that, for all these constructions,
complications can ensue when there are non-Higgsable clusters
\cite{MorrisonTaylorClusters}, which correspond to an anti-canonical divisor
$-K$ in the F-theory base geometry that contains a rigid component, giving
additional gauge factors that may intersect the divisors carrying the $\SM$
gauge group. The clearest situations in which models can be understood under
these circumstances are those in which the divisors $b_3$, $b_2$, and $\beta$
%(or $-4K-2b_3-b_2$)
for class \labelcref{item:beta>=0}
%(or \labelcref{item:Y=0})
models do not contain or intersect any of the divisors carrying non-Higgsable
clusters,  so that the additional gauge and matter factors act like decoupled
dark matter sectors, even after unHiggsing to the enhanced group. This simple
kind of circumstance only arises for models of class \labelcref{item:beta>=0},
however, since the condition $Y = 0$ implies that any divisor supporting a
non-Higgsable gauge component must be contained in one of the divisors $b_3,
b_2, -4 K - 2 b_3 - b_2$ associated with the gauge factors of the unHiggsed
Pati--Salam model. For a more detailed analysis of these issues see
\cite{RaghuramTaylorTurner321W}.

%%%%%%%%%%%%%%%%%%%%%%%%%%%%%%%%%%%%%%%%%%%%%%%%%%%%%%%%%%%%%%%%%%%%%%%%%%%%%%
%%%%%%%%%%%%%%%%%%%%%%%%%%%%%%%%%%%%%%%%%%%%%%%%%%%%%%%%%%%%%%%%%%%%%%%%%%%%%%
\subsection{4D F-theory models}\label{sec:F-theory-4D}

We can now generalize the F-theory constructions of class
\labelcref{item:beta>=0} and class \labelcref{item:Y=0} models described for 6D
theories to 4D F-theory. For 4D models, the anomaly coefficient products for
localized matter fields appearing in \cref{tab:generic} are replaced with the
curve given by the intersection of the corresponding divisor classes, so that
these entries in the table describe the matter curve supporting matter in the
appropriate representation. (Note that the same replacement does not apply for
the adjoint representations, as these are not supported locally on curves.)

For 4D models, the geometry of the solutions of class \labelcref{item:Y=0} are
again constructed using the fiber $F_{11}$ as described in
\cite{KleversEtAlToric}.  For the models of class \labelcref{item:beta>=0}, the
starting geometry will again be a deformation of the $\SMuh$ Tate model
geometry.  Even without an explicit Weierstrass description, the geometry of
the matter curves is still encoded in \cref{tab:generic}.

Just as, for example, the Tate model describes generic F-theory constructions
with an $\SU(N)$ gauge group and generic matter in 4D as well as in 6D, it is
natural to hypothesize that the two classes of F-theory constructions
identified here will also give the general F-theory constructions of models
with gauge group $\SM$ and the generic matter representations listed in
\cref{tab:generic} in both 6D and 4D.

The class of models recently studied in \cite{CveticEtAlQuadrillion} correspond
to the 4D models of class \labelcref{item:Y=0} with the specific choice $b_3 =
b_2 = -K$ over a weak Fano base.  Note that, as mentioned above, this also
corresponds to the class of models that are reached by Higgsing an $\SU(5)$
gauge theory tuned on the divisor $-K$. The divisor classes associated with the
gauge factors are the same in these models, and allow for gauge coupling
unification.  On the other hand, for most models in class
\labelcref{item:beta>=0}, the divisors supporting the three gauge factors are
distinct and independent, so we do not expect gauge coupling unification in
those models.

As for six-dimensional models, all the constructions described here can be
carried out over any weak Fano base, and one can in principle try to tune a
Standard Model gauge group in class \labelcref{item:beta>=0} over any complex
threefold base that supports an elliptic Calabi--Yau fourfold. Recent studies
suggest that the number of these base geometries is enormous, on the order of
$10^{3000}$ \cite{HalversonLongSungAlg,TaylorWangLandscape}, though for larger
bases the prevalence of many non-Higgsable clusters limits the number of
divisors available for tuning additional Standard Model structure.

Also note that, since in 4D the chiral matter content is controlled by fluxes,
even though there are matter curves for various representations not appearing
in the Standard Model in these constructions, particularly for those of class
\labelcref{item:beta>=0}, this does not mean that these matter representations
must appear as chiral matter in any particular corresponding supersymmetric
F-theory model. In fact, 4D anomaly cancellation will impose fairly strong
restrictions on what chiral matter combinations are possible from fluxes (as
in, e.g., \cite{CveticGrimmKleversAC}). On the other hand, it is interesting to
note that the models of class \labelcref{item:Y=0} with $b_2, b_3 < -a$ are the
only ones that contain only the matter multiplets of the MSSM.

As an example of 4D $\SM$ models with generic matter over a given base, we can
consider the base $\bP^3$.  A simple calculation shows that there are 181
models in class \labelcref{item:beta>=0} with divisors satisfying the
constraints \labelcref{eq:X-constraint,eq:Y-constraint}, with $a = K = -4 H$,
where $H$ is the hyperplane class on $\bP^3$.  There are 54 models in class
\labelcref{item:Y=0}, of which 6 are also in class
\labelcref{item:beta>=0}.

%%%%%%%%%%%%%%%%%%%%%%%%%%%%%%%%%%%%%%%%%%%%%%%%%%%%%%%%%%%%%%%%%%%%%%%%%%%%%%
%%%%%%%%%%%%%%%%%%%%%%%%%%%%%%%%%%%%%%%%%%%%%%%%%%%%%%%%%%%%%%%%%%%%%%%%%%%%%%
%%%%%%%%%%%%%%%%%%%%%%%%%%%%%%%%%%%%%%%%%%%%%%%%%%%%%%%%%%%%%%%%%%%%%%%%%%%%%%
\section{Outlook}\label{sec:outlook}

We have shown that in 6D supergravity without tensor multiplets, two simple
classes of F-theory models combine to give all anomaly-free models with gauge
group $\SM$ and the generic matter multiplets listed in \cref{tab:generic}.  By
analogy with other known F-theory constructions, it is natural to conjecture
that these two classes of models will give the generic constructions of both 4D
and 6D models with geometrically tuned gauge group $\SM$.

One of the most striking features of this analysis is that we have found that
the largest set of F-theory constructions that share the Standard Model gauge
group and matter representations live in a three-parameter family of models
that come from a deformation of $\SMuh$ theory.  If this is true in four
dimensions as well as in six dimensions with minimal supersymmetry, it is
possible that this structure may even be relevant in analyzing Standard Model
constructions from string theory in vacua without supersymmetry.

This analysis also highlights the significance of the second class of models
encountered, which corresponds to the $F_{11}$ fiber constructions analyzed in
\cite{KleversEtAlToric}.  While the first class \labelcref{item:beta>=0} gives
a larger, three-parameter family of models and is compatible with a wider
range of bases including those with possible (supersymmetrically) non-Higgsable
hidden dark matter sectors, the second class \labelcref{item:Y=0} contains
fewer potentially nonzero matter multiplicities for non-MSSM matter and many
models in this class have a more explicit description through toric geometry.

The focus of this note has been on the underlying geometry of generic F-theory
constructions of the Standard Model. While in six dimensions, this geometry
essentially completely determines the physics, for four dimensional models the
geometry is just a starting point; fluxes must be included to produce chiral
matter, and many other effects must be considered \cite{WeigandTASI}. The
classes of models described in this note with the Standard Model gauge group
and matter content would be promising to explore further for realistic
four-dimensional physics models from F-theory, keeping in mind that all the
models described here essentially have a completely tuned gauge group, and are
disjoint from other possible constructions that use non-Higgsable components in
the gauge group or rely on flux breaking of a GUT.

\subsection*{Acknowledgments}
We would like to thank Mirjam Cveti\v{c}, Jim Halverson, Elias Kiritsis, Ling
Lin, Paul Oehlmann, Massimo Porrati, Fernando Quevedo, Nikhil Raghuram, Gary
Shiu, and Yinan Wang for helpful discussions. The authors are supported by DOE
grant DE-SC00012567.

%%%%%%%%%%%%%%%%%%%%%%%%%%%%%%%%%%%%%%%%%%%%%%%%%%%%%%%%%%%%%%%%%%%%%%%%%%%%%%
%%%%%%%%%%%%%%%%%%%%%%%%%%%%%%%%%%%%%%%%%%%%%%%%%%%%%%%%%%%%%%%%%%%%%%%%%%%%%%
%%%%%%%%%%%%%%%%%%%%%%%%%%%%%%%%%%%%%%%%%%%%%%%%%%%%%%%%%%%%%%%%%%%%%%%%%%%%%%
%%%%%%%%%%%%%%%%%%%%%%%%%%%%%%%%%%%%%%%%%%%%%%%%%%%%%%%%%%%%%%%%%%%%%%%%%%%%%%
%\appendix

%%%%%%%%%%%%%%%%%%%%%%%%%%%%%%%%%%%%%%%%%%%%%%%%%%%%%%%%%%%%%%%%%%%%%%%%%%%%%%
%%%%%%%%%%%%%%%%%%%%%%%%%%%%%%%%%%%%%%%%%%%%%%%%%%%%%%%%%%%%%%%%%%%%%%%%%%%%%%
%%%%%%%%%%%%%%%%%%%%%%%%%%%%%%%%%%%%%%%%%%%%%%%%%%%%%%%%%%%%%%%%%%%%%%%%%%%%%%

\bibliographystyle{JHEP}
\bibliographystyle{plain}
\bibliography{research}

\end{document}